\documentclass[preprint,aps,prb,superscriptaddress]{revtex4-2}

\usepackage{graphicx}
\usepackage{multirow}
\usepackage{bm}
\usepackage{color}
\usepackage{amsmath}
\usepackage{ulem}
\usepackage{soul}
\usepackage{balance}
\usepackage{lineno}
\usepackage[pdfstartview=FitH, CJKbookmarks=true, bookmarksnumbered=true, bookmarksopen=true, colorlinks, pdfborder=001, linkcolor=blue, anchorcolor=blue, citecolor=blue,urlcolor=blue,breaklinks]{hyperref}

\usepackage{amssymb}

\begin{document}

\title[Xiaoying Zheng, et al]{Coherent magnetic excitations in a topological Kondo semimetal
}

\author{Xiaoying Zheng}
\affiliation{Center for Correlated Matter and School of Physics, Zhejiang University, Hangzhou 310058, China}

\author{Devashibhai T. Adroja}
\affiliation{ISIS Facility, STFC Rutherford Appleton Laboratory, Harwell Oxford, Oxfordshire OX11 0QX, United Kingdom}
\affiliation  {Highly Correlated Matter Research Group, Physics Department, University of Johannesburg, P.O. Box 524, Auckland Park 2006, South Africa}

\author{Hiroaki Kadowaki}
\affiliation{Department of Physics, Tokyo Metropolitan University, Hachioji, Tokyo 192-0397, Japan}

\author{Rajesh Sharma}
\affiliation{Department of Physics, Indian Institute of Science, Bangalore-560012, India}

\author{Tanmoy Das}
\affiliation{Department of Physics, Indian Institute of Science, Bangalore-560012, India}

\author{Seiko Ohira-Kawamura}
\affiliation{Neutron Science Section, MLF, J-PARC Center, Shirakata, Tokai, Ibaraki 319-1195, Japan}

\author{Maiko Kofu}
\affiliation{Neutron Science Section, MLF, J-PARC Center, Shirakata, Tokai, Ibaraki 319-1195, Japan}

\author{Zhaoyang Shan}
\affiliation{Center for Correlated Matter and School of Physics, Zhejiang University, Hangzhou 310058, China}

\author{Toshiro Takabatake}
\affiliation{Department of Quantum Matter, AdSE, Hiroshima University, Higashi-Hiroshima 739-8530, Japan}
\affiliation{Center for Correlated Matter and School of Physics, Zhejiang University, Hangzhou 310058, China}

\author{Huiqiu Yuan}
\affiliation{Center for Correlated Matter and School of Physics, Zhejiang University, Hangzhou 310058, China}
\affiliation{Institute for Advanced Study in Physics, Zhejiang University, Hangzhou 310058, China.}
\affiliation{Institute of Fundamental and Transdisciplinary Research, Zhejiang University, Hangzhou 310058, China.}
\affiliation{State Key Laboratory of Silicon and Advanced Semiconductor Materials, Zhejiang University, Hangzhou 310058, China.}

\author{Chao Cao}
\affiliation{Center for Correlated Matter and School of Physics, Zhejiang University, Hangzhou 310058, China}

\author{Michael Smidman}
\affiliation{Center for Correlated Matter and School of Physics, Zhejiang University, Hangzhou 310058, China}

\date{\today}

\begin{abstract}
	\textbf{In Kondo insulators the many-body Kondo lattice effect drives the formation of bands containing heavy charge carriers with a hybridization gap, leading to insulating properties. These renormalized bands can host non-trivial topologies driven by strong electron-electron interactions, but probing narrow heavy bands at low temperatures is challenging. We use inelastic neutron scattering (INS) to probe the  Kondo lattice CeNiSn, which hosts both semimetallic transport properties and a hybridization gap. The INS response exhibits momentum-dependent magnetic excitations and a spin-gap in the low-temperature Kondo coherent state, which electronic structure calculations corroborate as arising from the renormalized heavy band structure. Dynamical-mean field theory demonstrates that this renormalized band structure corresponds to a topological Kondo insulating state, and hence the INS probes bulk excitations of heavy topological bands. This identification of a Kondo insulator addresses the long-standing mystery of the electronic properties of CeNiSn, and demonstrates the manifestation of a topological many-body coherent state in spectroscopic measurements of strongly correlated narrow band materials.}
\end{abstract}
\maketitle

Strongly correlated electron systems are characterized by the presence of macroscopic phenomena arising due to many body electron-electron interactions that cannot be accounted for by one-electron physics. Kondo lattices are exemplary examples of such systems in which there is Kondo-hybridization between the lattice of  $f$-electrons and the conduction electron sea, and coherence between the periodically arranged localized electrons gives rise to the formation of a narrow band with significant $f$-character at low temperatures \cite{Coleman2006}. The characteristic features of the renormalized electronic structure near the Fermi level $E_{\rm F}$ are a greatly enhanced density of states together with the opening of a hybridization gap. Typically this gives rise to a Fermi-liquid state with heavy quasiparticle masses, but if $E_{\rm F}$ lies within the hybridization gap then a Kondo insulating state forms at low temperatures \cite{Riseborough2000}.

Moreover, the strong spin-orbit coupling and different parity between $f$-electron (odd parity) and conduction electron (even parity) states means that there can be a band-inversion in the low temperature Kondo insulating state \cite{Dzero2016}. SmB$_6$ is proposed to be an example of such a topological Kondo insulator \cite{Takimoto2011}, where evidence has been found for topological surface states within the bulk hybridization gap from both angle-resolved photoemission spectroscopy (ARPES) \cite{Xu2013,Neupane2013,Jiang2013,Xu2014} and scanning tunnelling spectroscopy \cite{Ruan2014,Rossler2014,Jiao2018}, although the topological nature and origin of these states are debated \cite{Zhu2013,Hlawenka2018}. The topological Kondo insulating state also manifests in transport measurements where the increase of resistivity upon cooling arising from the opening of an insulating gap in the bulk is truncated by a resistivity plateau dominated by surface conduction \cite{Wolgast2013}. SmB$_6$ has a sizeable bulk hybridization gap of around 20~meV that forms at elevated temperatures. However, in Kondo lattice systems with small Kondo energy scales $T_{\rm K}$ of a few Kelvin, the coherent $f$-bands are similarly  narrower and form at correspondingly lower temperatures, making it challenging to probe the topological renormalized electronic structure with ARPES. On the other hand, inelastic neutron scattering (INS) has been shown to be a powerful probe of the heavy coherent bands of Kondo lattice systems, where measurements with higher energy resolutions and lower temperatures are readily realized \cite{Fuhrman2015,Goremychkin2018}.

In the heavy fermion CeNiSn there is  a different manifestation of the  low temperature Kondo coherent state in the  transport properties, whereby resistivity measurements indicate the opening of a Kondo hybridization gap, but the system never reaches the full Kondo insulating state \cite{Takabatake1990,Takabatake1992}, and the best samples were eventually shown to be metallic at low temperatures \cite{Izawa1999,Terashima2002}. This peculiar behavior is also reflected by the evidence for a partial or highly anisotropic gap from  nuclear magnetic resonance \cite{Kyogaku1990,Nakamura1994} and tunneling spectroscopy \cite{Ekino1995}, that was generally ascribed to a highly anisotropic or nodal Kondo hybridization \cite{Stockert2016,Ikeda1996,Moreno2000,Macedo2025}. More recently the role of band topology in the coherent Kondo state of CeNiSn has come into question, where it was proposed that the non-symmorphic symmetry gives rise to topologically protected M{\"o}bius-twisted surface states \cite{Chang2017,Yoshida2019}, while ARPES measurements of the high temperature band structure suggest that the uncorrelated state corresponds to a Dirac nodal-loop semimetal with hourglass band crossings \cite{Nam2019,Bareille2019}. However, both experimental and theoretical probes of the nature of the low temperature renormalized coherent Kondo state of CeNiSn are still lacking.

Inelastic neutron scattering  measurements of CeNiSn pose another long-standing puzzle, in that at low temperatures magnetic excitations accompanied by a spin-gap are only observed for a limited range of momentum transfers, namely a 2~meV excitation is found at (0~1~0) and (0~0~1) \cite{Mason1992,Sato1995}, as well as quasi-one-dimensional 4~meV excitations at ($Q_a~K+1/2~Q_c$) \cite{Kadowaki1994,Sato1995}. Given the lack of magnetic ordering in CeNiSn  \cite{Kratzer1992}, together with the onset of the magnetic excitations corresponding to the Kondo coherence energy scale, a natural suggestion was that these excitations were related to the transitions between the heavy renormalized coherent $f$-bands \cite{Ikeda1996,Raymond1997}, but these could also be related to  short-range antiferromagnetic correlations \cite{Kadowaki1994,Sato1995} or a heavy spin-liquid driven by hybridization of the crystal field states with the conduction electrons \cite{Kagan1993}. The aforementioned magnetic excitations were revealed through INS measurements with a triple-axis spectrometer, which can only cover a relatively restricted region of energy and momentum transfers, while the development in recent years of high-flux cold-neutron time-of-flight spectrometers allows for low energy magnetic excitations to be mapped across an extended region of four-dimensional $\mathbf{Q}-\omega$ space. Here we report INS measurements of single crystals of CeNiSn using such a spectrometer that reveal  extended momentum-dependent low energy magnetic excitations, which indeed correspond to transitions of topological heavy bands in the coherent Kondo state.

\begin{figure}[tb]
	\includegraphics[scale=0.95]{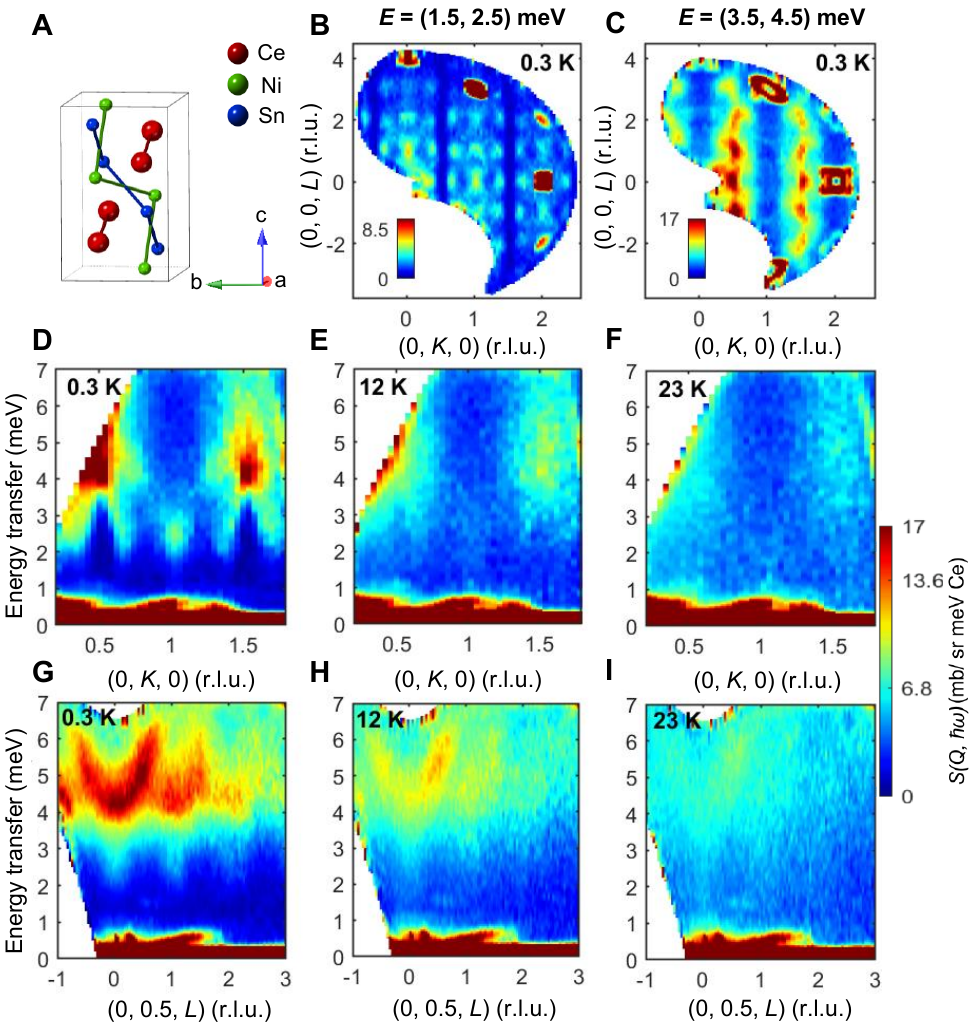}%
	\caption{Crystal structure and inelastic neutron scattering measurements of CeNiSn. (\textbf{A}) Crystal structure of CeNiSn, where red, green and blue correspond to Ce, Ni and Sn atoms respectively. Constant energy slices of inelastic neutron scattering measurements with an incident energy of $E_i=8.97$~meV at 0.3~K in the (0$KL$) scattering plane are displayed integrating over energy transfers of (\textbf{B}) 1.5 - 2.5~meV, and (\textbf{C}) 3.5 - 4.5~meV. Color plots of the excitation spectra are shown at three temperatures for momentum transfers of  (\textbf{D})-(\textbf{F}) (0,~$K$~0), and  (\textbf{G})-(\textbf{I}) (0,~0.5~$L$).}
\label{Fig1}
\end{figure}

\begin{figure}[tb]
	\includegraphics[scale=0.47]{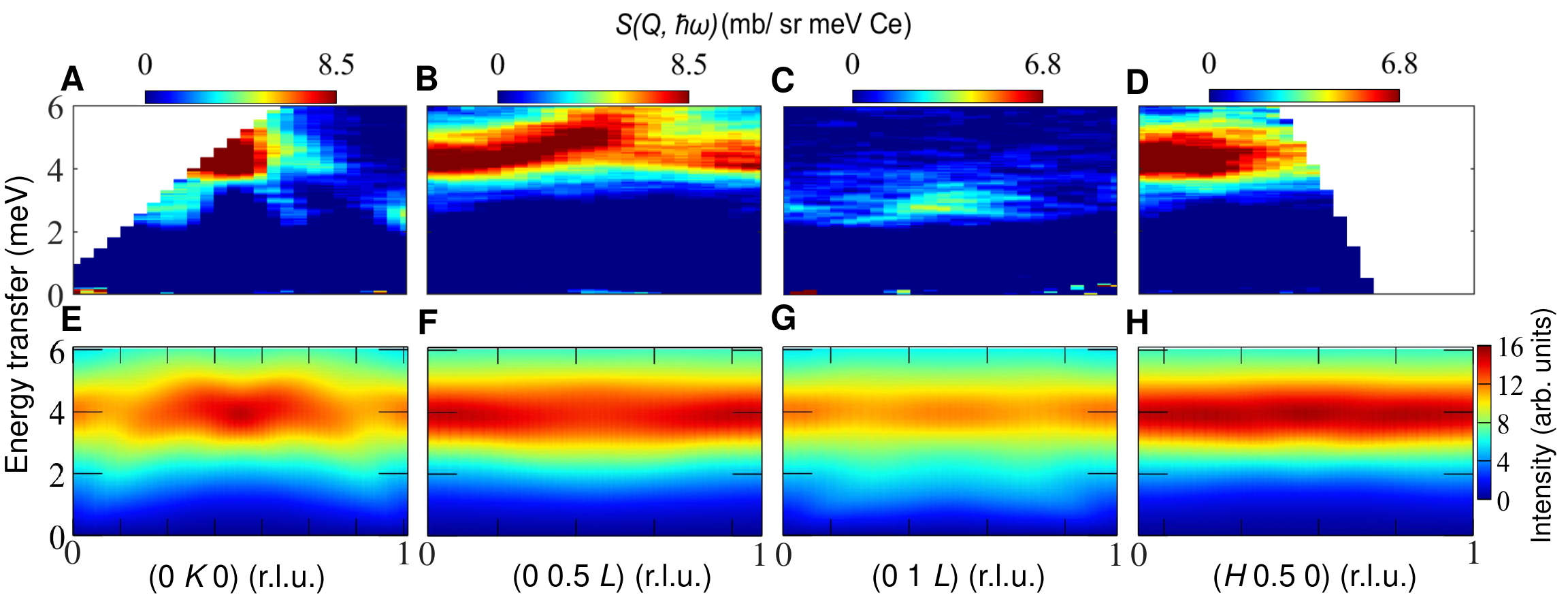}%
	\caption{Low temperature magnetic excitations of CeNiSn. Estimates of the intensity of the magnetic excitations of CeNiSn $S(\mathbf{Q},\hbar\omega)$ at 0.3~K in units of (mb/sr~meV~Ce), obtained by subtracting the 23~K data from that at 0.3~K, along momentum transfers (\textbf{A}) (0~$K$~0), (\textbf{B}) (0~0.5~$L$), (\textbf{C}) (0~1~$L$), and (\textbf{D}) ($H$~0.5~0). (\textbf{E})-(\textbf{H})  show the corresponding imaginary part of the dynamical susceptibility calculated using linear response theory based on DFT+$U$ calculations  within the random-phase approximation (Eq.~\ref{eq-chi_rpa}).}
\label{Fig3}
\end{figure}

\begin{figure}[tb]
	\includegraphics[scale=0.6]{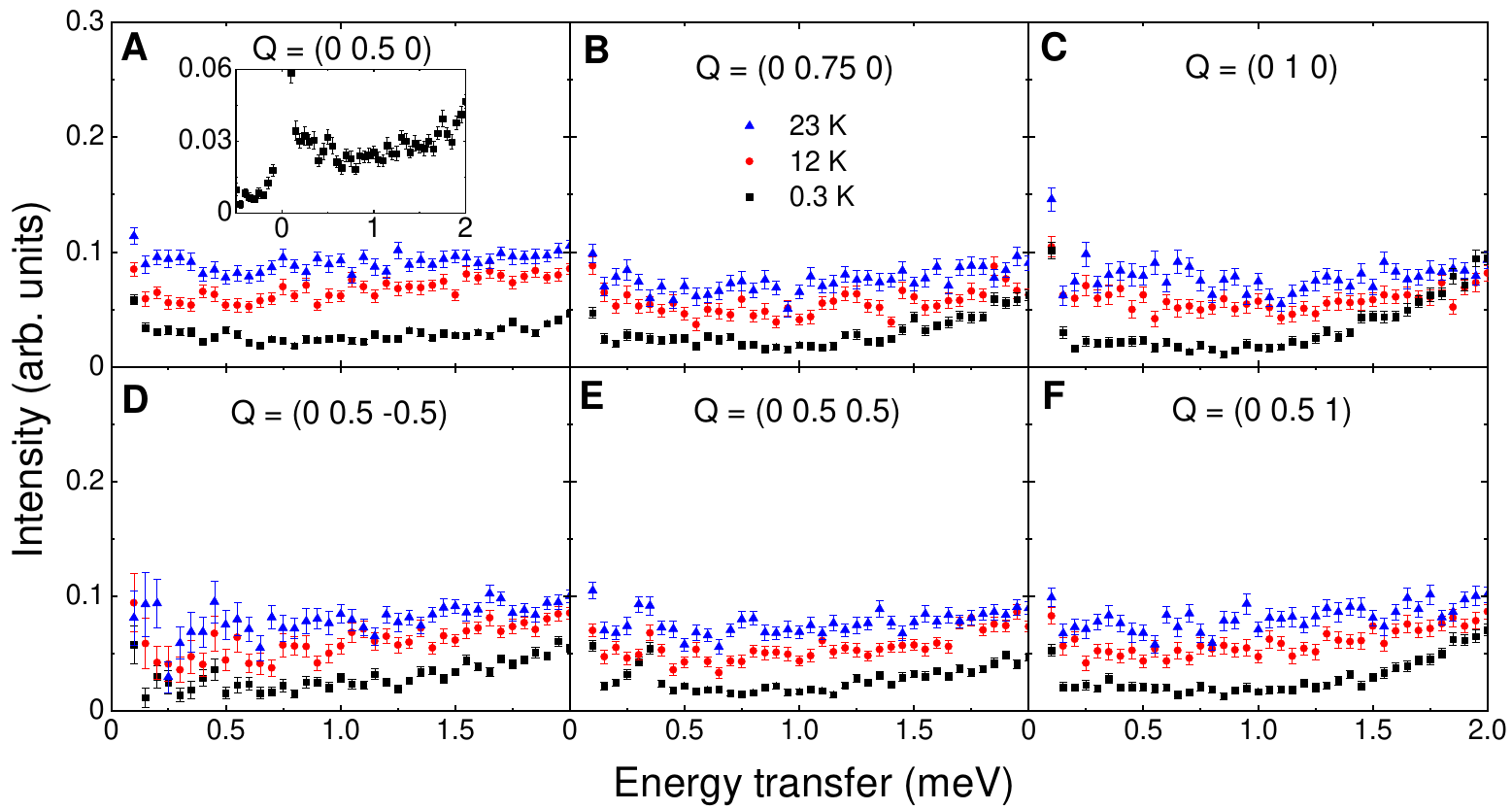}%
	\caption{Low energy cuts of inelastic neutron scattering measurements of CeNiSn. Low energy cuts of measurements with an incident energy of $E_i=3.44$~meV as a function of energy transfer are compared for the three measured temperatures for different momentum transfers $\mathbf{Q}$, with integration widths of $\Delta H=\Delta L= \pm0.2$~r.l.u, $\Delta K=\pm0.1$~r.l.u for (\textbf{A})-(\textbf{C})  and  $\Delta H=\Delta K= \pm0.2$~r.l.u, $\Delta L=\pm0.1$~r.l.u for (\textbf{D})-(\textbf{F}) .  The reduced intensity at low energies at 0.3~K evidences the opening of an extended spin-gap. The inset of panel a highlights the quasielastic scattering present at 0.3 K.}
\label{Fig2}
\end{figure}

\noindent\textbf{Inelastic neutron scattering on CeNiSn}

Figures~\ref{Fig1}B and C display constant energy slices of INS measurements at 0.3~K on single crystals of CeNiSn in the $(0KL)$ scattering plane with an incident energy $E_i=8.97$~meV. With an energy integration range of 1.5 - 2.5~meV, the  magnetic excitations with the strongest intensity are centered at moment transfers ($\mathbf{Q}$) of (0 1 0) and (0 0 1), which correspond to the previously observed 2~meV excitation \cite{Mason1992,Kadowaki1994,Sato1995}. There is also additional magnetic scattering in this energy range for non-integer $K$, situated either side of $K=0.5$. At higher energies in the range 3.5 - 4.5~meV, strong magnetic scattering is observed at $K=0.5$ for all $Q_c$, in line with the previously observed quasi-one-dimensional 4~meV excitation along $\mathbf{Q_b}$ \cite{Kadowaki1994,Sato1995}. Meanwhile at intermediate energy ranges (2.5 - 3.5~meV; Fig. S2), ring-shaped magnetic scattering is observed (note that the strong ring features around integer ($H K L$) in Figs.~\ref{Fig1}B and C are from phonon scattering). Together  these show that the aforementioned magnetic scattering constitute extended coherent magnetic excitations in CeNiSn.

The momentum-dependence of the magnetic excitations at 0.3~K can be seen in Figs.~\ref{Fig1}D and G, which show  the INS spectra along (0~$K$~0) and (0~0.5~$L$), respectively. Along (0~$K$~0), the strongest intensity is at around 4~meV at the half-integer $K$, i.e. (0~0.5~0) and (0~1.5~0), and the energy of the excitations decreases moving away from these momenta. Meanwhile along (0~0.5~$L$), the excitations initially move upwards in energy with increasing $L$ that reaches a maximum around $L=0.7$. At 12~K the $\mathbf{Q}$-dependence of the magnetic excitations is not readily detected, but there is still appreciable magnetic scattering at $K=0.5$, which is weaker still at 23~K. In order to more clearly resolve the $\mathbf{Q}$-dependence of the low temperature magnetic excitations, the data measured at 23~K were subtracted from that at 0.3~K, which is shown for four directions in Figs.~\ref{Fig3}A-D. Peak positions of constant-$\mathbf{Q}$ cuts were also extracted (Figs. S4 and S5). Together these show that along (0~$K$~0) the excitation has maximum energy and intensity at (0 0.5 0), which moves to lower energies reaching a minimum of around 2~meV at (0~1~0), while there is a second weaker branch that decreases more rapidly to a similar energy.   Meanwhile the excitation along (0~0.5~$L$) splits into two branches, while there is little discernable momentum dependence along ($H$~0.5~0).

At 0.3~K, it can also be seen that the scattering intensity is decreased at low energies below the coherent magnetic excitations, signifying the opening of a spin-gap. One-dimensional cuts integrated over different momentum transfers are displayed in Fig.~\ref{Fig2} for $E_i=3.44$~meV, which all show a depletion of the intensity at low energies at 0.3~K compared to 12 and 23~K, suggesting that the spin-gap opens over extended regions of the Brillouin zone (although this is less pronounced for $\mathbf{Q}\parallel\mathbf{a^*}$; Fig. S3), rather than only being present in a narrow momentum range \cite{Mason1992,Kadowaki1994,Sato1995}. On the other hand, the persistence of a quasielastic contribution at the lowest temperature shown in the inset of Fig.~\ref{Fig2}A is consistent with the presence of in-gap states, as inferred previously \cite{RAYMOND1998245}.

\noindent\textbf{Comparison with DFT + U and RPA}

In order to understand the origin of the momentum-dependent magnetic excitations, electronic structure calculations were performed using the DFT+$U$ method (Fig. S6), in which the bands at the Fermi level $E_F$ primarily consist of Ce-$4f$ orbitals, with some contribution from Ce-$d$ orbitals. To compare the calculations with the experimental data, the imaginary part of the dynamical susceptibility was calculated using linear response theory within the random-phase approximation (RPA) (Eq.~\ref{eq-chi_rpa}), which are shown in Figs.~\ref{Fig3}E-H taking into account an effective mass renormalization of a factor of 60, as commonly considered for heavy fermion compounds \cite{Song2021}. It can be seen that this calculated dynamical susceptibility can reproduce certain features of the 0.3~K data, namely there is a maximum intensity at (0 0.5 0), which decreases in energy together with a weaker intensity towards the zone center. At the zone center there is an increase in intensity, in line with the 2~meV scattering at these momentum transfers in the data. There is also additional intensity at lower energies close to (0 0.5 0), which is also consistent with the lower branch observed in the data.  Moreover, the calculated dynamical  susceptibility lacks momentum-dependence along (0 1 $L$) and ($H$ 0.5 0), in agreement with the data.  On the other hand, the data shows a stronger momentum dependence than in the calculations, and the 2~meV  excitation close to (0.~0.5~1) is less well reproduced. Similarly the calculations also show the slight upturn observed between (0 0.5 0) and (0 0.5 0.5), but the splitting at higher $L$ is not resolved. Since these calculations exhibit  many of the salient aspects of the data, this suggests that the  magnetic excitations at low temperature correspond to transitions related to the heavy renormalized band structure, although some of the detailed features are not reproduced, which can be a consequence of the  DFT+$U$ calculations being inadequate to describe the many-body strongly correlated Kondo-lattice ground state.

\noindent\textbf{DMFT calculations and topological renormalized band structure}

\begin{figure*}[htp]
  \includegraphics[width=16cm]{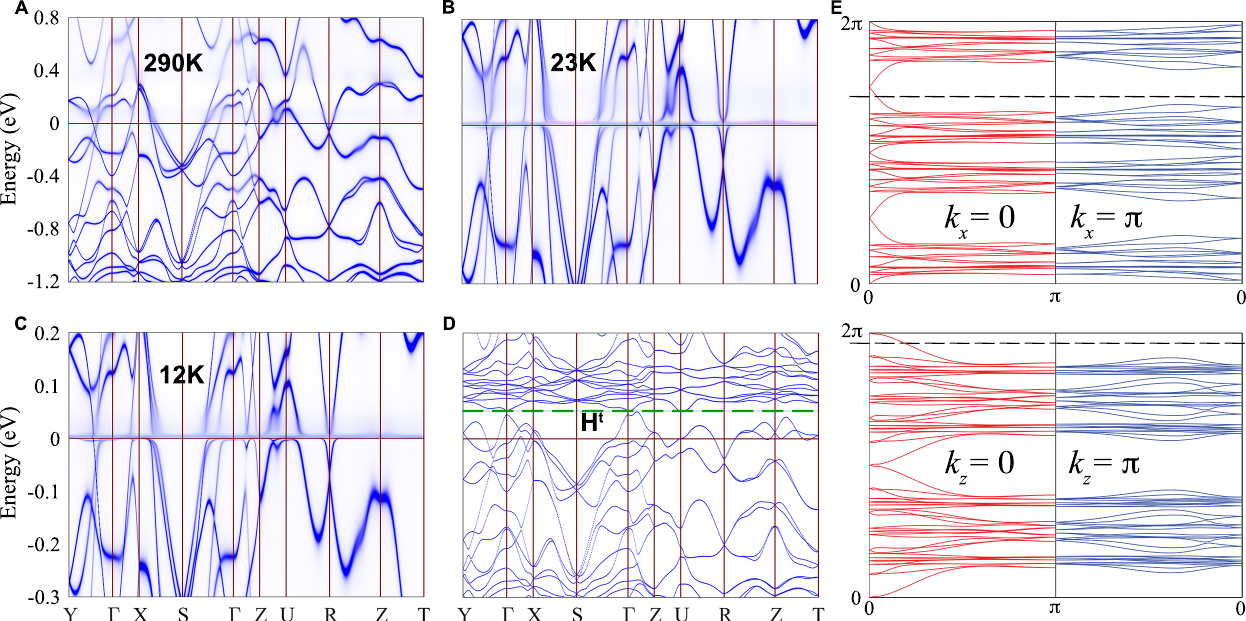}
  \caption{\label{fig:dmft}Momentum resolved spectral function of CeNiSn from dynamical mean-field theory. The calculated spectral functions are displayed for (\textbf{A}) 290~K, (\textbf{B}) 23~K,  (\textbf{C})12~K. (\textbf{D}) Eigenstates of the topological Hamiltonian $H^t$. The dashed line intersects the fully open direct gap between the $N$-th and $(N+1)$-th states. (\textbf{E}) The Wannier charge center (WCC) evolution calculated using $H^t$ at $k_i=0$ $k_i=\pi$ ($i$=x or z) planes. The WCC evolution line intersects with the dashed line odd/even times at the $k_i=0$/$\pi$ planes respectively, indicating a strong topological insulator. }
\end{figure*}

In order to gain further insights into the renormalized electronic structure and its temperature evolution, we have performed DFT+DMFT calculations for CeNiSn at 290~K, 23~K and 12~K. At high temperatures (290 K), the Ce-4$f$ orbitals form local moments, resulting in  incoherent $f$-states (Fig. \ref{fig:dmft}A). The resulting momentum resolved spectral function is therefore very similar to the DFT calculations assuming a Ce$^{3+}$ state without 4$f$ valence states. As the temperature decreases, the Ce-4$f$ states accumulate weight close to $E_F$, as the hybridization between 4$f$ states and conduction electrons becomes stronger. In addition, the 4$f$ states gradually becomes coherent, and sharp quasiparticle states form (Fig. \ref{fig:dmft}B-C). As a result, significant band-bending appears near $E_F$, and a hybridization gap starts to emerge. However, the temperature is still above the Kondo-coherence temperature, which is estimated to be around 9~K from the peak in the Hall coefficient \cite{Takabatake1992b}, and this could  account for the opening of a full gap not being observed in our calculations. The energy and momentum dependence of the  imaginary part of the dynamical susceptibility is also calculated under RPA using the particle-hole susceptibility derived from the DFT+DMFT calculations at 12~K, which is shown in Fig. S7. These also show similar features to that obtained from DFT+$U$ in Figs.~\ref{Fig3}E-H, where the strongest intensity being at around 7~meV, close to that observed in the INS data, demonstrates that the strong band renormalization can largely be accounted for by ab-initio calculations.

The nature of the renormalized band structure can still be examined, by projecting the 12~K results onto the topological Hamiltonian $H^t=H_0+\Sigma(\omega=0)$ (Fig.~\ref{fig:dmft}D), where $H_0$ is the non-interacting Hamiltonian and $\Sigma(\omega=0)$ is the zero-frequency self-energy, which reflects the band topology at $E_{\rm F}$ \cite{PhysRevX.2.031008}. This exhibits a gap between the $N$-th and $(N+1)$-th states across the whole Brillouin zone, where $N$ is the number of electrons, consistent with a fully open direct gap. However, it can be seen that the top of the valence band near $\Gamma$, and the bottom of the conduction band near $\mathbf{U}$ are very similar in energy, in line with a closed indirect gap. Figure~\ref{fig:dmft}E shows the results from Wilson loop calculations, where the topological invariant is given by the winding number. In the $k_i=0$ planes, the Wannier charge centers cross the reference line once, indicating that the $\mathcal{Z}_2$ invariant is 1 for these planes, whereas in the $k_i=\pi$ planes, they do not cross the reference line, so the $\mathcal{Z}_2$ invariant is 0. Thus, the complete $\mathcal{Z}_2$ invariant is (1;000), and therefore the renormalized electronic structure indeed corresponds to a topological Kondo insulating state.

\noindent\textbf{Discussion}

Comparison between the low temperature INS spectra and RPA calculations suggest that the magnetic excitations can be understood as arising from the heavy renormalized band structure that emerges at low temperatures from the lattice Kondo effect. Moreover, the projection of DMFT calculations  onto the  topological Hamiltonian shows that this Kondo coherent state is a topological insulating state, and hence the magnetic excitations correspond to interband transitions of topological renormalized bands. In fact, the momentum dependence of the eigenvalues of this topological Hamiltonian is quite similar to the band dispersions from DFT+$U$ calculations, which can be identified from comparisons around the dashed line in Fig.~\ref{fig:dmft}D with Fig. S6.  Note that  the dashed line in Fig.~\ref{fig:dmft}D  separates the $N$-th and $(N+1)$-th states and therefore it is expected to coincide with the Fermi level at sufficiently low temperature.  The DFT+$U$ results in Fig. S6 do not have spin-orbit coupling included in the calculations, and thus the band crossings due to inversion between $f_{x(x^2-3y^2)}$ and $f_{xyz}$ lead to metallic states between $G-X$ and $G-Z$. Such band crossings are lifted when spin-orbit coupling is considered, leading to a $\mathcal{Z}_2$ topological state. This is consistent with the conclusion from DMFT calculations that  this Kondo coherent state has a direct gap over the entire Brillouin zone with a $\mathcal{Z}_2$ invariant. The DMFT calculations at 12~K still lack the full opening of the gap, which is in line with the INS measurements at this temperature showing a reduced depletion of the low energy spectral weight compared to 0.3~K.

A hallmark of a topological insulating state are gapless topological surface states whose presence is ensured due to different topology between the bulk band structure and the vacuum. While INS is only sensitive to the bulk excitations, the identification of a topological Kondo insulating state in CeNiSn suggests the need to search for novel topological surface states \cite{Chang2017,Yoshida2019} using surface sensitive probes such as ARPES or scanning tunneling microscopy. Although the presence of Dirac surface states is guaranteed in the single-particle picture, for the renormalized band structure arising from electronic correlations, such states may not be realized due to the evolution of the many body ground state upon approaching the surface. For instance, the anticipated surface states of the Weyl heavy fermion semimetal YbPtBi \cite{Guo2018} are only observed in ARPES measurements of Yb-terminated surfaces \cite{Fang2023}, likely due to a valence transition in the outermost Yb-ions which pushes the surface states out of the bulk states. Moreover in SmB$_6$ the putative topological surface states appear to emerge well below the temperatures at which the Kondo insulating gap opens, which has been ascribed to either a suppressed surface Kondo effect \cite{Jiao2016}, or lack of gap coherence at elevated temperatures \cite{Pirie2020}. Evidence for multiple  energy scales is also reported for CeNiSn, including the formation of states inside the pseudogap at low temperatures \cite{Izawa1999}, and therefore combined bulk and surface probes are necessary to disentangle their origin and the relationship to the evolution of the correlated topological state.

~\\
\noindent\large{\textbf{Methods}}
~\\
\noindent\normalsize{\textbf{Crystal growth and inelastic neutron scattering}}

Single crystals of CeNiSn were grown by the Czocharalski method using the $^{58}$Ni isotope so as to reduce the incoherent neutron scattering cross section. Two single crystals (total mass 10.74 g) were coaligned for the INS experiments, which were performed using the AMATERAS cold-neutron time-of-flight spectrometer at the J-PARC facility. The sample was orientated with the $a$-axis vertical, such that $bc$ corresponds to the horizontal scattering plane, and spectra were measured at different orientations upon rotating through 135$^{\circ}$ in the scattering plane. Measurements were performed using multi-$E_i$ mode with incident energies (energy resolution) $E_i$ = 3.44(0.067), 5.25(0.123), 8.97(0.27), 18.7(0.8)~meV. The momentum transfer $\mathbf{Q}=(H, K, L)$ corresponds to reciprocal lattice units, $\mathbf{Q}=(q_xa/2\pi, q_yb/2\pi, q_zc/2\pi)$, where $a=7.542$, $b=4.601$, and $c=7.617\AA$, are the orthorhombic cell lattice parameters \cite{TAKABATAKE1994457}. The intensities were normalized to absolute units by applying a scaling factor obtained from a comparison to previous INS results (Fig. S1) \cite{Sato1995}.

\noindent\normalsize{\textbf{DFT+U calculations}}

The electronic structure of CeNiSn was investigated using density-functional theory (DFT) simulations implemented in the Vienna Ab initio Simulation Package (VASP) \cite{vasp1}. To model the interactions between electrons, the generalized gradient approximation (GGA) parameterized by Perdew-Burke-Ernzerhof (PBE) was employed \cite{pbe}. The interactions between core and valence electrons were characterized using the projector-augmented wave (PAW) approach \cite{paw1}, using a orthorhombic unit cell (space group $Pnma$ (No. 62)) with lattice parameters  $a$, $b$ and $c$ of 7.545 ,  4.605 and  7.627 \AA, respectively.  For band diagram and Fermi surface calculations,  the plane-wave energy was set with a cut-off to 520 eV, and  the convergence threshold for the electronic self-consistent cycle was established at $10^{-6}$ eV. We used the simplified method suggested by Dudarev et al. \cite{Dudarev1998} to perform the PBE$+$U calculations, considering the difference between $U$ and $J$. Specifically, we assigned $U$ values of 7.0 eV for Ce(4$f$), 5.1 eV for Ni(3$d$), and 3.5 eV for Sn(3$d$) \cite{aflowlib}, while maintaining $J$ at 0.7 eV for Ce(4$f$) and zero for other atoms.

\noindent\normalsize{\textbf{Multiband RPA susceptibility}}

We present first-principles multi-band spin susceptibility calculations within the random-phase approximation (RPA) for CeNiSn. We consider a multi-band Hubbard model given by
\begin{equation}
H_{int}=\sum_{i,\sigma}U_i n_{i,\sigma}n_{i,\bar{\sigma}}+\sum_{i \neq j,\sigma,\sigma'}V_{i,j}n_{i,\sigma}n_{j,\sigma'} 
\label{eq-Hubbard}
\end{equation}
Here, $U_i$ is the intraband Hubbard interaction in band $i$ and $ V_{i,j}$ is the inter-band Hubbard interaction between the bands $i$ and $j$, and these Hubbard interaction terms are reorganized into charge channels $\tilde{U}_c$ and spin channels $\tilde{U}_s$.

To compute the RPA spin and charge susceptibilities, $\chi_{s/c}$, we directly incorporate the density functional theory (DFT) band structures.
$\tilde{\chi}_{s/c}$ are the density-density correlators for the spin and charge density channels. We define the noninteracting density-density correlation function (Lindhard susceptibility) $\tilde{\chi_0}$ within the standard linear response theory :
\begin{equation}
\begin{aligned}
\left[ \chi_0(\textbf{q},\omega) \right]_{i,j}^{l,m} &= -\dfrac{1}{\Omega^2_{BZ}}\sum_{\textbf{k},\nu,\nu'}
\phi_j^{\nu}(\textbf{k}) \phi_i^{\nu\dagger}(\textbf{k}) \phi_m^{\nu'}(\textbf{k}+\textbf{q}) \phi_l^{\nu'\dagger}(\textbf{k}+\textbf{q}) \\
& \times \dfrac{f(\bar{E}_{\nu'}(\textbf{k}+\textbf{q}))-f(\bar{E}_{\nu}(\textbf{k}))}{\omega+\bar{E}_{\nu'}(\textbf{k}+\textbf{q})-\bar{E}_{\nu}(\textbf{k})+i\epsilon}
\end{aligned}
\label{eq-chi0}
\end{equation}

Many body effects of Coulomb interaction in the density- density correlation are captured within the $S$-matrix expansion of the Hubbard Hamiltonian in Eq. (\ref{eq-Hubbard}). By summing over different bubble and ladder diagrams, the RPA spin and charge susceptibilities are obtained as:
\begin{equation}
\tilde{\chi}_{s/c}(\textbf{q},\omega)=\tilde{\chi}_{0}(\textbf{q},\omega)\left(\tilde{\mathbb{I}}\mp\tilde{U}_{s/c}\tilde{\chi}_0(\textbf{q},\omega)\right)^{-1}
\label{eq-chi_rpa}
\end{equation}
where $\tilde{\mathbb{I}}$ is the unit matrix.

The self-energy correction to the electronic structure is introduced within the Fermi-liquid ansatz by renormalizing the band structures $\bar{E}_{\nu}({\bf k}) = ZE_{\nu}({\bf k})$ using a band-independent renormalization factor $Z=1/60$, which is comparable to the DMFT result near the Fermi level.

\noindent\normalsize{\textbf{DFT+DMFT calculations}}

For the DFT+DMFT calculations, we have employed $U$=6.0 eV and $J$=0.7 eV (corresponding to $F^0$=6.0 eV, $F^2$=8.345 eV, $F^4$=5.5747 eV, $F^6$=4.1226 eV) for the Ce-4$f$ orbitals. The first Brillouin zone is sampled with 2600 K-points ($11\times19\times11$). The states within [-10.0, 10.0] eV with respect to the Fermi level are projected to obtain the low-energy Hamiltonian employed in the DMFT calculations. The spin-orbit coupling is considered as a second-order perturbation, and the crystal-field splitting (CEF) is considered at the lattice level. The continuous time quantum Monte Carlo (CT-QMC) impurity solver was employed, with the full Coulomb interaction matrix, and the 4$f$ impurity occupations were constrained within [0, 3]. A nominal double counting scheme with $n_f^0=1$ was chosen.

In addition,  the minimum position of the imaginary part of the self-energy around $\omega=0$ was identified by solving:
$$\frac{\partial \mathrm{Im}[\Sigma(\omega)]}{\partial \omega}\vert \omega_0=0$$
We note that at low temperatures, $\omega_0$ is close to 0, and the corresponding $\mathrm{Im}[\Sigma(\omega_0)]$ is also negligible. At 12K, $\omega_0\approx 0.2$ meV, and $\mathrm{Im}[\Sigma_{j=5/2}(\omega_0)]\approx-0.014$ eV. 

By taking the quasiparticle approximation, the self-energy $\Sigma(\omega)$ can be expanded to  first order around $\omega_0$ using:
$$\Sigma(\omega)=\Sigma(\omega_0)+(\omega-\omega_0)\frac{\partial \Sigma(\omega_0)}{\partial \omega}$$
The quasiparticle weight is defined as:
 $$Z=\left(1-\frac{\partial \Sigma(\omega_0)}{\partial \omega}\right)^{-1}$$
In the above equations, $\frac{\partial \Sigma(\omega_0)}{\partial \omega}$ is short for $\frac{\partial \Sigma(\omega)}{\partial \omega}\vert_{\omega_0}$. By definition, it is real quantity. Under such an approximation, the Green's function is:
$$G^{qp}_{\mathbf{k}}(\omega)=Z^{1/2}\left[\omega-H^{qp}\right]^{-1}Z^{1/2}$$
with a quasiparticle Hamiltonian.
$$H^{qp}=Z^{1/2}\left(H^0_{\mathbf{k}}+\Sigma_0-\omega_0(1-Z^{-1})-H^{dc}\right)Z^{1/2}$$
where $H^{dc}$ is the double-counting term. The dispersion of $H^{qp}$ is shown overlaid with the calculated spectral function in Fig. S8.

Under the quasiparticle approximation, the bare particle-hole susceptibility can be calculated using:
$$\chi^{qp}_{\mathbf{q}}(i\nu)=-\frac{1}{N_{\mathbf{k}}\beta}\sum_{n\mathbf{k}} G^{qp}_{\mathbf{k+q}}(i\omega_n+i\nu) G^{qp}_{\mathbf{k}}(i\omega_n)$$
The above calculation was performed on a $24\times36\times24$ K-mesh and imaginary frequency axis.

Using the above bare susceptibility, the general susceptibility matrix is then obtained from the Bethe-Salpeter equation:
$$\chi_{\mathbf{q}}(i\nu)=\chi^{qp}_{\mathbf{q}}(i\nu)+\chi^{qp}_{\mathbf{q}}(i\nu)V\chi_{\mathbf{q}}(i\nu)$$
where $V$ is the full interaction matrix calculated using the given $U$ and $J$. The spin and charge susceptibilities are then obtained by properly tracing $\chi_{\mathbf{q}}(i\nu)$. The real-axis quantities are eventually obtained through analytic continuation using Pad\'{e} method, and the calculated imaginary part of the spin susceptibility is shown in Fig. S7.

\noindent\normalsize{\textbf{Topological Hamiltonian and Wilson loop method}}

The topological Hamiltonian~\cite{PhysRevX.2.031008} is defined as 
$$H^t_{\mathbf{k}}=H^0_{\mathbf{k}}+\Sigma(\omega=0)-H^{dc}$$
where $\Sigma(\omega=0)$ is the self-energy at zero frequency, and $H^{dc}$ is the double counting term. By definition, the topological Hamiltonian reproduces the Fermi surface of the DMFT Hamiltonian. In addition, it yields the same generalized Chern number of the original interacting Hamiltonian, which reduces to the conventional Thouless, Kohmoto, Nightingale, den Nijs (TKNN) invariant in the non-interacting limit and is directly related to the quantum anomalous Hall coefficient. For the topological Hamiltonian, the Wilson loop method, which tracks the evolution of the Wannier charge centers along certain directions in a 2D manifold in reciprocal space, can be applied to obtain the Chern number.

~\\
\noindent\large{\textbf{Acknowledgments}}
~\\
\noindent\normalsize{}
This work was supported by the National Key R$\&$D Program of China (Grants No. 2022YFA1402200, and No. 2023YFA1406303) and the National Natural Science Foundation of China (Grants No. 12222410, No. 12274364, No. 11974306, No. 12034017, and No. 12174332). D.T.A. would like to thank the Royal Society of London for International Exchange funding between the UK and Japan, Newton Advanced Fellowship funding between UK and China, the CAS for PIFI Fellowship, and EPSRC UK for the funding (Grant No. EP/W00562X/1).	R.S. acknowledges the Science and Engineering Research Board (SERB), Government of India, for providing the NPDF fellowship with grant number PDF/2021/000546. T.D. acknowledges funding from Core Research Grant (CRG) of S.E.R.B.  (CRG/2022/003412-G) and benefited from the computational funding from the National Supercomputing Mission (DST/NSM/RDHPC$-$Applications/2021/39), both are under the Department of Science and Technology, India. The neutron experiment at the Materials and Life Science Experimental Facility of the J-PARC was performed under a user program (Proposal No. 2020A0080).

~\\
\noindent\large{\textbf{Data Availability}}
~\\
\noindent\normalsize{}
The datasets generated during the current study are available from the corresponding author on reasonable request.

~\\
\noindent\large{\textbf{Additional Information}}
~\\
\noindent\normalsize{}
Correspondence and requests for materials should be addressed to D. T. Adroja (devashibhai.adroja@stfc.ac.uk), T. Das (tnmydas@gmail.com), C. Cao (ccao@zju.edu.cn), or M. Smidman (msmidman@zju.edu.cn).

~\\
\noindent\large{\textbf{Author Contributions}}
~\\
\noindent\normalsize{}
The project was conceived by D.T.A. and M.S.. The crystals were grown by T.T., and inelastic neutron scattering measurements were performed by H.K., S.O.K. and M.K.. The data were analyzed by X.Z., D.T.A., H.K., S.O.K., M.K., Z.S., H.Y. and M.S.. DFT + U based calculations were performed by R.S. and T.D., and DFT+DMFT based calculations were performed by C.C.. The manuscript was written by X.Z., R.S., T.D., C.C., and M.S., with input from all authors.

\end{document}